# Giant enhancement in the thermal responsivity of microelectromechanical resonators by internal mode coupling

Ya Zhang[1,a)], Ryoka Kondo[2], Boqi Qiu[2], Xin Liu[4], and Kazuhiko Hirakawa[2,3,b)]

[1]*Institute of Engineering, Tokyo University of Agriculture and Technology, Koganei-shi, Tokyo, 184-8588, Japan*

[2] *Institute of Industrial Science, University of Tokyo, 4-6-1 Komaba, Meguro-ku, Tokyo 153-8505, Japan*

[3]*Institute for Nano Quantum Information Electronics, University of Tokyo, 4-6-1 Komaba, Meguro-ku, Tokyo 153-8505, Japan*

[4] *Beijing-Dublin International College & Institute of Theoretical Physics, Beijing University of Technology, 100 Pingleyuan, Beijing, 100124, P.R. China*

We report on a giant enhancement in the thermal responsivity of the doubly-clamped GaAs microelectromechanical (MEMS) beam resonators by using the internal mode coupling effect. This is achieved by coupling the fundamental bending mode with the fundamental torsional mode of the MEMS beam resonators through the cubic Duffing nonlinearity. In the mode coupling regime, we have found that, when the input heat to the MEMS resonators is modulated at a particular frequency, the resonance frequency shift caused by heating can be enhanced by almost two orders of magnitude. The observed effect is promising for realizing high-sensitivity thermal sensing by using MEMS resonators, such as ultrasensitive terahertz detection at room temperature.

---

a) Electronic mail: zhangya@go.tuat.ac.jp

b) Electronic mail: hirakawa@iis.u-tokyo.ac.jp

Microelectromechanical (MEMS) resonators[1-3] are very attractive for sensing applications owing to their intrinsic high sensitives. The high mechanical quality (Q)-factors of the MEMS resonators enable the detection of small changes in the resonance frequency, which can be used for detecting changes in mass,[4-6] charge,[7,8] spin orientation,[9-11] temperature,[12,13] and infrared radiation.[14] A resonant MEMS sensor is often described by a spring-mass model. The sensing target such as molecules, charges, temperature, etc., modulates either the spring constant or the mass of the system and the signal is detected as a shift in the resonance frequency. The responsivity is simply determined by how much the sensing target modulates the spring constant or the mass. Such a simple operation principle makes it difficult to dramatically improve the responsivity of the MEMS sensors.

The mode coupling effect in the MEMS resonators have attracted considerable interest since it changes the properties of the MEMS resonators substantially. A vibrational mode is coupled with a higher mode through the internal resonance, when their frequencies fulfill the integer ratio of 1 : $N$, where $N$ is the nonlinear order of the MEMS resonator[15-19]. In the mode coupling regime, the energy is transferred from the externally driven mode to other modes. The use of such an effect was proposed for realizing frequency stabilization[15] and synchronization[20], vibrational energy harvesting[21], energy dissipation control[22,23], and the detection of higher resonance modes[24]. The mode coupling effect is also very attractive for sensing applications owing to its rich features in oscillation amplitude and frequency of the MEMS resonators. Indeed, it was used for increasing the responsivities in the mass sensing applications[25,26], in which a steady state feature of the internal mode resonance was employed. However, dynamical effects of the internal mode resonance on MEMS sensors have been much less explored[27].

Here, we report on giant enhancement in the thermal responsivity of the doubly-clamped GaAs MEMS beam resonators[28-30] by using the internal mode coupling effect. This is achieved by coupling the fundamental bending mode with the fundamental torsional mode of the MEMS beam resonators through the cubic Duffing nonlinearity. In the mode coupling regime, we have found that, when the input heat to the MEMS resonators is modulated at a particular frequency, the resonance frequency shift caused by heating can be enhanced by almost two orders of magnitude. The observed effect is promising for realizing high-sensitivity thermal sensing by using MEMS resonators, such as ultrasensitive terahertz detection at room temperature, and also has a potential for other sensing applications.

We fabricated GaAs doubly clamped MEMS beam resonators by using a modulation-doped AlGaAs/GaAs heterojunction grown by molecular beam epitaxy.[31] After growing a 200-nm-thick GaAs buffer layer and a 3-μm-thick $Al_{0.7}Ga_{0.3}As$ sacrificial layer on a (100)-oriented semi-insulating GaAs substrate, the beam layer was formed by depositing a 1-μm-thick GaAs layer. We subsequently grew a 20-nm-thick Si-doped GaAs layer, a 70-nm-thick $Al_{0.3}Ga_{0.7}As$ layer and a 10-nm-thick GaAs capping layer. The suspended beam structure was formed by selectively etching the sacrificial layer with diluted hydrofluoric acid. On both ends of the beam, we fabricated an ohmic contact and a surface Schottky gate to form piezoelectric capacitors for driving the beam.[28,30,31] An *ac* voltage was applied to one of the piezoelectric capacitors to drive the beam and the induced resonant beam motion was monitored by a laser Doppler vibrometer. The resonance signal was fed to a PLL to provide a feedback control for maintaining a self-oscillation and read the resonance amplitude and frequency by using the built-in lock-in detection function of the PLL. On the MEMS beam, we deposited a 10-nm-thick NiCr layer, whose sheet resistance was ~1000 Ω/□, as a heater for thermally modulate the resonance frequencies and also calibrate the responsivity of the MEMS resonator. All the measurements were performed in a vacuum (~$10^{-4}$ torr) at room temperature.

To induce the internal mode coupling via the cubic Duffing nonlinearity in doubly clamped MEMS beam resonators, we need two resonance modes whose resonance frequencies fulfill the relationship $f_1: f_2 = 1:3$. Here we employ the fundamental bending mode ($f_b$) and the fundamental torsional mode ($f_t$) for building the internal mode coupling. The ratio ($f_b:f_t$) can be controlled, because $f_b$ is mainly determined by the length of the MEMS beam, whereas $f_t$ is mainly determined by the width. We designed the doubly clamped MEMS beam resonators with a geometry of 128(*L*)×30(*W*)×1.2(*t*) μm³ in such a way that $f_b$ is slightly smaller than 1/3 of $f_t$. We applied an ac driving voltage $V_D = 20$ mV to a MEMS beam resonator (sample A) and swept the driving frequency. The resonance spectrum was measured by using a phase locked loop (PLL) with a built-in function of the lock-in detection, as schematically shown in Fig. 1(b). The PLL was also used to provide a feedback control for maintaining a self-oscillation, as we reported elsewhere[30]. Figure 1(b) plots the measured spectrum for the first three modes, *i. e.*, the first bending mode, the second bending mode, and the first torsional mode. Note that $f_b = \sim 313.8$ kHz and $f_t = 958.0$ kHz. The inset of Fig. 1(b) shows the mode shapes of the first bending mode and the first torsional mode calculated by the finite element method. Figure 1(c) shows the measured resonance spectra of the first bending mode at various driving voltages ($V_D = 10\text{-}50$ mV) applied to the piezoelectric capacitor[31]. The Q-factor of the resonance was about 5,000. When the first bending mode is driven into the nonlinear Duffing oscillation regime ($V_D > 20$ mV), $f_b$ increases with increasing driving amplitude, which is due to the hardening effect of the GaAs MEMS beam. By using this effect, we tuned $f_b$ to make it approach $f_t/3$ ~320 kHz.

Figure 1(d) shows the measured resonance spectra when $V_D$ was increased to 200 mV-400 mV. When $f_b$ is increased to about 320 kHz, a small reduction in the resonance amplitude is observed in the spectrum, indicating that the vibrational energy of the fundamental mode decreases in this condition. Figure 1(e) shows a blow-up of the spectrum in the mode coupling region marked by a red dotted rectangle in Fig.1(d). Two clear drops in the amplitude appear at ~320.6 kHz and ~320.9 kHz, suggesting that the internal mode coupling is formed between the driven fundamental mode and a higher mode. Since the internal mode coupling in the MEMS beam resonators usually arises from the cubic-term nonlinearity, we identify the higher mode in this mode coupling to be the first torsional mode ($f_t = 958.0$ kHz). When the two modes couple with each other, they are renormalized into two new eigenmodes, $f_L$ and $f_H$ ($f_L$: the lower frequency coupled mode, $f_H$: the higher frequency coupled mode). These two new coupled modes appear as two dips in the resonance spectrum of the $f_b$ mode at around 320 kHz ($f_L$~320.6 kHz and $f_H$~320.9 kHz) as shown in Fig. 1(e), since the vibrational energy is transferred to the $f_t$ mode at these frequencies. Note that the frequency difference between $f_L$ and $f_H$ ($\Delta f$) was ~300 Hz for sample A. The internal mode coupling behavior was also observed in another sample (sample B) and the frequency difference $\Delta f$ was ~500 Hz (Supplementary Information I).

To study the effect of the internal mode coupling to the thermal responsivity of the MEMS resonator, we applied a dc voltage to the NiCr heater to generate a heat on the MEMS beam and measured $f_b$ as a function of the heating power, $P$. Figure 2 shows $f_b$ for sample A as a function of $P$ at various $V_D$ (283-424mV). The red and black curves show the data when $P$ is swept from 0 to 0.25 mW (forward) and from 0.25 mW to 0 (backward), respectively. In general, $f_b$ is smoothly reduced when $P$ is increased due to thermal expansion of the MEMS beam, which has been used for standard thermal sensing.[28,30] When $P$ is swept backward, $f_b$ shows a plateau at $f_1$ = 320.6 kHz, which corresponds to the first amplitude drop in Fig. 1(e). When $P$ is swept forward, on the other hand, a kink is observed at the frequency of the second amplitude drop ($f_2$ ~320.9 kHz) in Fig. 1(e), as indicated by a red dashed arrow. This behavior was explained by the interplay between the mechanical nonlinearity and the internal mode coupling effect, which was proposed to achieve the frequency stabilization in MEMS resonators[15]. Note that the frequency plateau makes the thermal responsivity vanishingly small.

Next, we applied a small modulated heat ($P$ ~ 25 nW) to the MEMS beam of sample A when the sample is operated in the mode coupling condition. Because of the frequency stabilization in the mode coupling region, the frequency shift of the MEMS bolometer is vanishingly small when the heat modulation frequency, $f_m$, is low (< 100 Hz). However, when the heat is modulated at a particular frequency (~300 Hz) under the internal mode coupling condition, we have observed a huge peak in the frequency shift, as shown by the red and blue curves in Fig. 3(a). For comparison, the black curve in Fig. 3(a) shows the frequency shift with the same heating power when the MEMS resonator is operated outside the mode coupling region. When $f_m$ ~ 300 Hz, the heat-induced frequency shifts in the mode-coupling region is 14 Hz ($V_D$ = 354 mV) and 20 Hz ($V_D$ = 424 mV), which are, respectively, 17.5 and 25.0 times higher than that (~0.8 Hz) outside the mode-coupling region ($V_D$ = 283 mV). The different frequency shifts at various driving voltages suggest that the enhancement factor can be modulated by controlling the mode coupling condition.

Figure 3(b) shows the shift in the resonance frequency for sample B measured inside and outside of the mode coupling region when $P$ = 20 nW is applied to the MEMS beam. When the MEMS resonator is operated outside of the mode coupling region ($V_D$= 650 mV), the frequency shift is about 0.7 Hz and almost does not depend on $f_m$. However, when the MEMS resonator is operated in the mode coupling region ($V_D$= 820 mV), the frequency shift shows a large peak over 40 Hz when $f_m \approx$ 530 Hz, which is more than 60 times higher than that in the uncoupled condition.

We have found that $f_m$ that induces a large thermal responsivity agrees well with the frequency difference, $\Delta f$, between the two dips in the resonance spectra ($f_L$ and $f_H$). This fact suggests that the greatly enhanced thermal responsivity originates from the coherent energy transfer between the two coupled modes. In order to understand the mechanism, we have measured how $f_b$ develops for the modulated heat input. Figure 4(b) plots the time trace of the shift in $f_b$ as a function of time when a small heat of ~20 nW(ac amplitude) at a modulation frequency of ~530 Hz is applied to the MEMS resonator. Figure 4(a) shows the waveform of the heat pulses fed to the MEMS beam. As seen, the frequency shift of $f_b$ is both positive and negative and gradually grows to a very large frequency shift. The time constant is in the order of ~100 ms. This is very different from the thermal response of the MEMS resonators without the mode coupling, where the resonance frequency only decreases due to thermal expansion when a heat is applied to the MEMS beam, as we reported before[30].

In the mode coupled condition, the MEMS resonator is driven at $f_L$, which corresponds to the first dip in Figs. 1(e). The pumped $f_L$-mode coherently transfer its vibrational energy to the $f_H$-mode through the parametric driving effect[32-35]. The parametrically excited $f_H$-mode forms a beating signal together with the $f_L$-mode, which periodically modulates the amplitude of the bending motions in the $f_L$- and $f_H$-modes, as schematically shown in Fig. 4(c). Note that we have bending motions at $f_L$ and $f_H$, and their constructive and destructive interferences modulate the amplitude of the bending motion at the frequency of $\Delta f = f_H - f_L$. Then, the $f_b$ is periodically modulated through the amplitude-frequency coupling that arises from the Duffing nonlinearity. Since the amplitude change of the bending motion by the interference is proportional to the amplitude of the $f_H$-mode, the periodic frequency shift becomes a sensor for measuring the amplitude of the $f_H$-mode[36]. Therefore, the enhanced frequency shift in Fig. 4(a) shows that the amplitude of the $f_H$-mode is increased by the parametric drive, until it becomes saturated due to the damping process.

Finally, let us discuss how the present mode coupling effect improves the detector performance. Figure 5(a) shows the frequency shift measured in the mode coupling region (450 Hz < $f_m$ < 600 Hz) for sample B, when $P$ is varied from 2 nW to 20 nW. Figure 5(b) plots the shift in the resonance frequency, $\Delta f$, determined from the curves plotted in Fig. 5(a) as a function of $P$. When $P \leq 10$ nW, $\Delta f$ increases almost linearly with increasing $P$. However, when $P > 10$ nW, $\Delta f$ gradually deviates from the linear increase. We have estimated the thermal responsivity, $R \equiv \Delta f/Pf_b \approx 10,000$ W$^{-1}$ by using a linear fitting to $\Delta f$ for $P \leq 10$ nW, as indicated by the dotted line in Fig. 5(b). The obtained $R$ in the mode coupling condition is almost two orders of magnitude larger than that in the uncoupled region.

In Fig. 5(c), we plot the frequency noise spectra, $n_f$, in (red) and outside (black) the mode coupling condition. As seen, at $f_m = 530$ Hz, where a large enhancement in the thermal responsivity is observed, $n_f$ also becomes larger ($n_f = \sim 50$ mHz/$\sqrt{\text{Hz}}$) when the MEMS resonator is operated under the mode coupling condition ($V_D = 820$ mV). Such enhancement in $n_f$ is not observed when the mode coupling is absent. Therefore, the noise equivalent power (NEP $\equiv n_f/Rf_b$) is not improved as much as $R$.

In thermal sensors, there exists a fundamental limit for NEP, namely, the thermal fluctuation limit, which is determined by the random transfer of thermal phonons between the MEMS beam and the thermal reservoir (the substrate in our device) and can be calculated as[37],

$$\text{NEP}_{\text{TF}} \approx (4k_B T^2 G_T)^{0.5} \approx 20 \text{ pW}/\sqrt{\text{Hz}} \quad \text{at } T = 300 \text{ K}, \tag{1}$$

where $k_B$ is the Boltzmann constant and $T$ the temperature. In Fig. 5(d), we plot NEP as a function of $f_m$ in (red) and outside (black) the mode coupling condition. When the resonator is operated in the mode coupling region ($V_D = 820$ mV), NEP becomes as small as 23 pW/$\sqrt{\text{Hz}}$ at $f_m = 530$ Hz. This NEP is close to the value of the thermal fluctuation limit, NEP$_{\text{TF}}$, which is indicated by a red dotted line in Fig. 5(d). When compared with NEP $\approx 150$ pW/$\sqrt{\text{Hz}}$ without the mode coupling ($V_D = 650$ mV), the improvement in NEP is about 6~7 times. Note that, although the mode coupling effect can improve $R$ by almost 2 orders of magnitude, the improvement in NEP is partly compensated by the increase in $n_f$. However, if extrinsic noise such as noise from amplifiers is dominant, the improvement in NEP will be governed by $R$.

In summary, we have demonstrated a giant enhancement in the thermal responsivity of doubly-clamped GaAs MEMS resonators using the internal mode coupling effect. The coupling between the fundamental bending mode and the fundamental torsional mode is formed by using the cubic Duffing nonlinearity in the system. When a static heat is applied to the MEMS beam, the frequency shift from the thermal effect is strongly suppressed, giving a vanishingly small thermal responsivity. However, when we apply a modulated heat to the MEMS beam at a particular modulation frequency of several hundreds of Hz, the thermally induced frequency shift is enhanced by almost two orders of magnitude, giving a giant enhancement in the thermal responsivity of the MEMS resonator. The enhancement is owing to the energy transfer between two renormalized coupled modes by parametric driving. The observed effect is promising for realizing high-sensitivity thermal sensing applications by using MEMS resonators, such as ultrasensitive terahertz detection at room temperature, and it also has a potential for other sensing applications.

We are grateful to Y. Arakawa and H. Yamaguchi for fruitful discussions. This work has been partly supported by JST Collaborative Research Based on Industrial Demand, MEXT Grant-in-Aid for Scientific Research on Innovative Areas "Science of hybrid quantum systems" (15H05868), and KAKENHI from JSPS (15K13966, 19K15023). Grants from the Precise Measurement Technology Promotion Foundation (PMTP-F) and the Murata Science Foundation are also acknowledged.

**Figure Captions**

**Figure 1** (a) Schematic illustration for the measurement setup. An *ac* voltage was applied to one of the piezoelectric capacitors to drive the beam and the induced resonant beam motion was monitored by a laser Doppler vibrometer. The motion signal was fed to a phase locked loop (PLL) circuit to provide a feedback control for maintaining the self-oscillation. The PLL was also used as a lock-in amplifier to measure the resonance spectrum. (b) The measured spectrum for the first three modes, *i. e.*, the first bending mode, the second bending mode, and the first torsional mode. The inset shows the mode shapes of the first bending mode and the first torsional mode calculated by the finite element method. (c) The measured resonance spectra of the first bending mode at various driving voltages ($V_D$ = 10-50 mV) applied to the piezoelectric capacitor. (d) The measured resonance spectra when $V_D$ was increased from 200 mV to 400 mV. (e) A blow-up of the spectrum in the mode coupling region marked by a dotted rectangle in Fig. 1(d). Two clear drops in the amplitude appear in the spectrum at ~320.6 kHz and ~320.9kHz, which are denoted as $f_L$ and $f_H$, respectively. $\Delta f$ is the difference between $f_L$ and $f_H$.

**Figure 2** The resonance frequency $f_b$ for sample A measured as a function of the input heating power P at various $V_D$ (283-424mV). The red and black curves show the data when *P* is swept from 0 to 0.25 mW (forward) and from 0.25 mW to 0 (backward), respectively. The red and black arrows indicate the sweep directions in the measurements. The red dashed arrow indicates the position of the kink in $f_b$.

**Figure 3** (a) Thermally induced frequency shift, $\delta f$, for sample A is plotted as a function of the modulation frequency of the applied heat (~25 nW) measured at various driving voltages, $V_D$. The black, blue and red curves plot the results when $V_D$= 283 mV, 354 mV and 424 mV, respectively. In the mode coupling region ($V_D$= 354 mV and 424 mV), the frequency shift is vanishingly small when the heat modulation frequency is low (<100 Hz), but shows a large peak when the heat is modulated at a particular frequency (~300 Hz). The black curve ($V_D$= 283 mV) show the frequency shift measured outside the mode coupling region. (b) Thermally induced $\delta f$ for sample B is plotted as a function of the modulation frequency of the applied heat (~20 nW) measured at various $V_D$. The black and red curves plot the results when $V_D$= 650 mV (outside the mode coupling region) and 820 mV (in the mode coupling region), respectively. $\delta f$ reaches its maximum when the heat is modulated at ~530 Hz.

**Figure 4** (a) Waveform of the input heat pulses fed to the MEMS beam. The applied heat power is ~20 nW (*ac* amplitude) and modulated at ~530 Hz. (b) Time trace of $\delta f$ measured when a small heat is applied to sample B. The MEMS resonator is operated in mode coupling regime ($V_D$ = 820 mV). (c) Schematic diagram of the beating effect between $f_L$- and $f_H$-modes. In the mode coupling condition, the bending mode and the torsional mode are renormalized into the new coupled modes at frequencies $f_L$ and $f_H$. Note that $f_L$- and $f_H$-modes consist of the bending component ($A_b$, $A'_b$) and the torsional component ($A_t$, $A'_t$). When the MEMS resonator is driven at $f_L$ by exciting the bending motion using the piezoelectric capacitor and, in addition, a periodic heat is applied to the beam at the frequency $\Delta f$ = $f_H$ - $f_L$, the $f_L$-mode coherently transfers its vibrational energy to the $f_H$-mode through the parametric driving effect. The parametrically excited $f_H$-mode generates a beating signal with the $f_L$-mode, which periodically modulates the amplitude of the bending component.

**Figure 5** (a) Thermally induced frequency shift in sample B measured in the mode coupling region when the heating power $P$ = 2, 4, 6, 8, 10, 16, and 20 nW and the heat modulation frequency, $f_m$, is swept from 450 Hz to 600 Hz. (b) Peak frequency shift $\Delta f$ in the mode coupling region is plotted as a function of $P$. The dotted line indicates a linear fit to $\Delta f$ for $P \leq 10$ nW to estimate the thermal responsivity. (c) Frequency noise spectra, $n_f$, measured when $V_D$= 650 mV (outside the mode coupling region; black) and 820 mV (in the mode coupling region; red). (d) NEP is plotted as a function of $f_m$ when $V_D$ = 650 mV (outside the mode coupling region; black) and 820 mV (in the mode coupling region; red). The red dotted line indicates the thermal fluctuation limit for NEP.

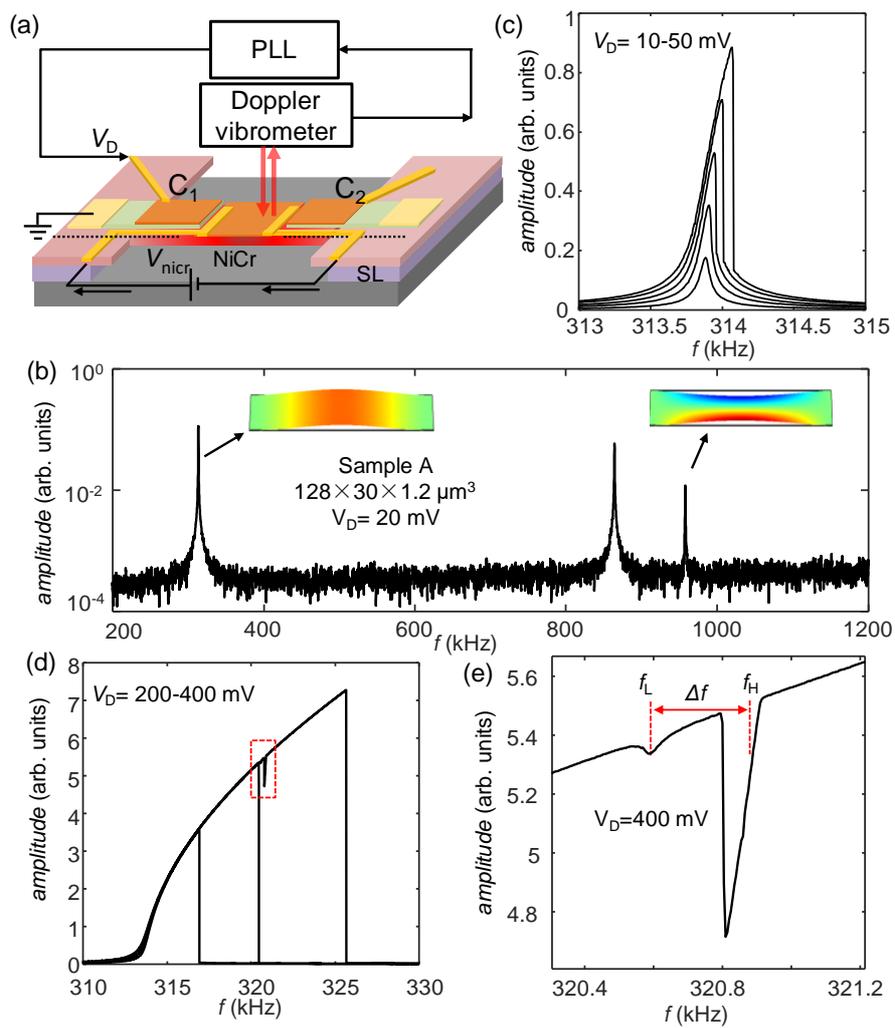

Fig. 1 Y. Zhang, et al.

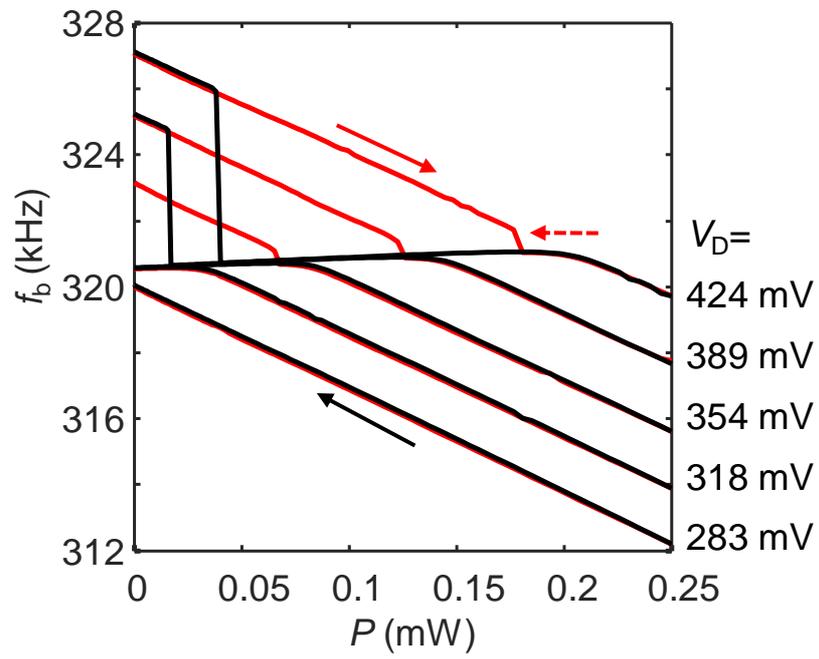

Fig. 2 Y. Zhang, et al.

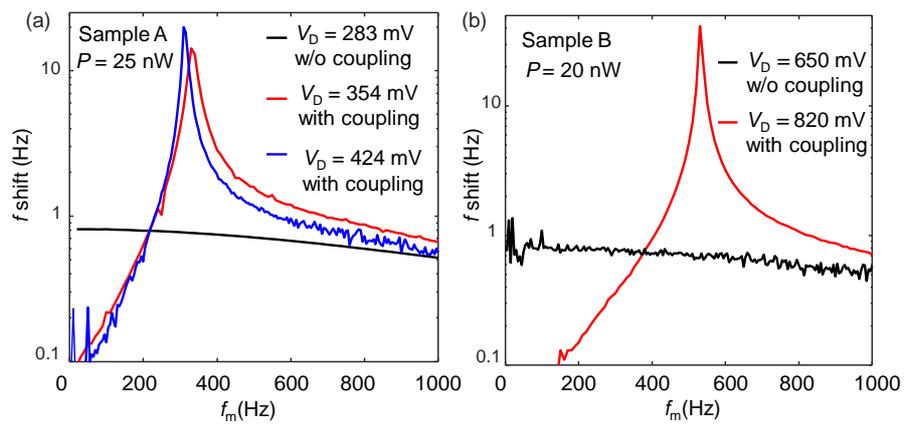

Fig. 3 Y. Zhang, et al.

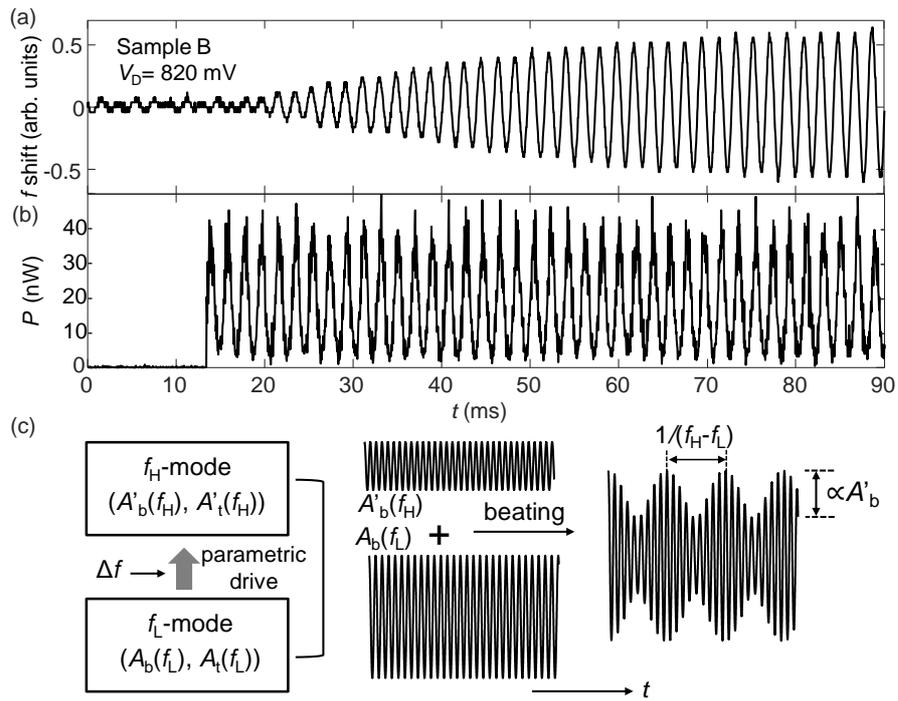

Fig.4  Y. Zhang, et al.

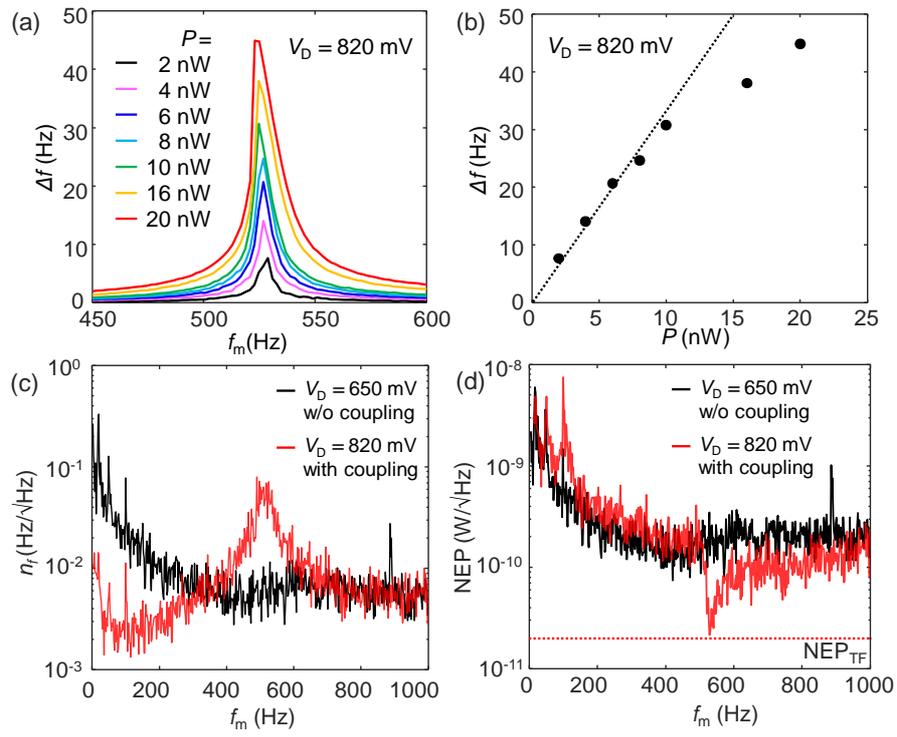

Fig. 5 Y. Zhang, et al.